\documentclass[amsmath,amssymb,aps,nofootinbib,nolongbibliography,superscriptaddress]{revtex4-2}

\usepackage{bm}
\usepackage{dcolumn}
\usepackage{graphicx}
\usepackage{multirow}
\usepackage{color}
\usepackage{mathtools}
\usepackage[T1]{fontenc}
\usepackage{lmodern}

\begin{document}

\title{Exploring atmospheric neutrino oscillations at ESSnuSB}

\author{J.~Aguilar}
\affiliation{Consorcio ESS-bilbao, Parque Cient\'{i}fico y Tecnol\'{o}gico de Bizkaia, Laida Bidea, Edificio 207-B, 48160 Derio, Bizkaia, Spain}
\author{M.~Anastasopoulos}
\affiliation{European Spallation Source, Box 176, SE-221 00 Lund, Sweden}
\author{E.~Baussan}
\affiliation{IPHC, Universit\'{e} de Strasbourg, CNRS/IN2P3, Strasbourg, France}
\author{A.K.~Bhattacharyya}
\affiliation{European Spallation Source, Box 176, SE-221 00 Lund, Sweden}
\author{A.~Bignami}
\affiliation{European Spallation Source, Box 176, SE-221 00 Lund, Sweden}
\author{M.~Blennow}
\affiliation{Department of Physics, School of Engineering Sciences, KTH Royal Institute of Technology, 106 91 Stockholm, Sweden}
\affiliation{The Oskar Klein Centre, AlbaNova University Center, Roslagstullsbacken 21, 106 91 Stockholm, Sweden}
\author{M.~Bogomilov}
\affiliation{Sofia University St. Kliment Ohridski, Faculty of Physics, 1164 Sofia, Bulgaria}
\author{B.~Bolling}
\affiliation{European Spallation Source, Box 176, SE-221 00 Lund, Sweden}
\author{E.~Bouquerel}
\affiliation{IPHC, Universit\'{e} de Strasbourg, CNRS/IN2P3, Strasbourg, France}
\author{F.~Bramati}
\affiliation{University of Milano-Bicocca and INFN Sez. di Milano-Bicocca, 20126 Milano, Italy}
\author{A.~Branca}
\affiliation{University of Milano-Bicocca and INFN Sez. di Milano-Bicocca, 20126 Milano, Italy}
\author{G.~Brunetti}
\affiliation{University of Milano-Bicocca and INFN Sez. di Milano-Bicocca, 20126 Milano, Italy}
\author{I.~Bustinduy}
\affiliation{Consorcio ESS-bilbao, Parque Cient\'{i}fico y Tecnol\'{o}gico de Bizkaia, Laida Bidea, Edificio 207-B, 48160 Derio, Bizkaia, Spain}
\author{C.J.~Carlile}
\affiliation{Department of Physics, Lund University, P.O Box 118, 221 00 Lund, Sweden}
\author{J.~Cederkall}
\affiliation{Department of Physics, Lund University, P.O Box 118, 221 00 Lund, Sweden}
\author{T.~W.~Choi}
\affiliation{Department of Physics and Astronomy, FREIA Division, Uppsala University, P.O. Box 516, 751 20 Uppsala, Sweden}
\author{S.~Choubey}
\email{Corresponding authors: S. Choubey, T. Ohlsson and S. Vihonen}
\affiliation{Department of Physics, School of Engineering Sciences, KTH Royal Institute of Technology, 106 91 Stockholm, Sweden}
\affiliation{The Oskar Klein Centre, AlbaNova University Center, Roslagstullsbacken 21, 106 91 Stockholm, Sweden}
\author{P.~Christiansen}
\affiliation{Department of Physics, Lund University, P.O Box 118, 221 00 Lund, Sweden}
\author{M.~Collins}
\affiliation{Faculty of Engineering, Lund University, P.O Box 118, 221 00 Lund, Sweden}
\affiliation{European Spallation Source, Box 176, SE-221 00 Lund, Sweden}
\author{E.~Cristaldo Morales}
\affiliation{University of Milano-Bicocca and INFN Sez. di Milano-Bicocca, 20126 Milano, Italy}
\author{P.~Cupia\l}
\affiliation{AGH University of Krakow, al. A. Mickiewicza 30, 30-059 Krakow, Poland }
\author{H.~Danared}
\affiliation{European Spallation Source, Box 176, SE-221 00 Lund, Sweden}
\author{J.~P.~A.~M.~de~Andr\'{e}}
\author{M.~Dracos}
\affiliation{IPHC, Universit\'{e} de Strasbourg, CNRS/IN2P3, Strasbourg, France}
\author{I.~Efthymiopoulos}
\affiliation{CERN, 1211 Geneva 23, Switzerland}
\author{T.~Ekel\"{o}f}
\affiliation{Department of Physics and Astronomy, FREIA Division, Uppsala University, P.O. Box 516, 751 20 Uppsala, Sweden}
\author{M.~Eshraqi}
\affiliation{European Spallation Source, Box 176, SE-221 00 Lund, Sweden}
\author{G.~Fanourakis}
\affiliation{Institute of Nuclear and Particle Physics, NCSR Demokritos, Neapoleos 27, 15341 Agia Paraskevi, Greece}
\author{A.~Farricker}
\affiliation{Cockroft Institute (A36), Liverpool University, Warrington WA4 4AD, UK}
\author{E.~Fasoula}
\affiliation{Department of Physics, Aristotle University of Thessaloniki, Thessaloniki, Greece}
\affiliation{Center for Interdisciplinary Research and Innovation (CIRI-AUTH), Thessaloniki, Greece}
\author{T.~Fukuda}
\affiliation{Institute for Advanced Research, Nagoya University, Nagoya 464–8601, Japan}
\author{N.~Gazis}
\affiliation{European Spallation Source, Box 176, SE-221 00 Lund, Sweden}
\author{Th.~Geralis}
\affiliation{Institute of Nuclear and Particle Physics, NCSR Demokritos, Neapoleos 27, 15341 Agia Paraskevi, Greece}
\author{M.~Ghosh}
\affiliation{Center of Excellence for Advanced Materials and Sensing Devices, Ruđer Bo\v{s}kovi\'c Institute, 10000 Zagreb, Croatia}
\author{A.~Giarnetti}
\affiliation{Dipartimento di Matematica e Fisica, Universit\'a di Roma Tre, Via della Vasca Navale 84, 00146 Rome, Italy}
\author{G.~Gokbulut}
\affiliation{University of Cukurova, Faculty of Science and Letters, Department of Physics, 01330 Adana, Turkey}
\affiliation{Department of Physics and Astronomy, Ghent University, Proeftuinstraat 86, B-9000 Ghent, Belgium}
\author{C.~Hagner}
\affiliation{Institute for Experimental Physics, Hamburg University, 22761 Hamburg, Germany}
\author{L.~Hali\'c}
\affiliation{Center of Excellence for Advanced Materials and Sensing Devices, Ruđer Bo\v{s}kovi\'c Institute, 10000 Zagreb, Croatia}
\author{M.~Hooft}
\affiliation{Department of Physics and Astronomy, Ghent University, Proeftuinstraat 86, B-9000 Ghent, Belgium}
\author{K.~E.~Iversen}
\affiliation{Department of Physics, Lund University, P.O Box 118, 221 00 Lund, Sweden}
\author{N.~Jachowicz}
\affiliation{Department of Physics and Astronomy, Ghent University, Proeftuinstraat 86, B-9000 Ghent, Belgium}
\author{M.~Jenssen}
\affiliation{European Spallation Source, Box 176, SE-221 00 Lund, Sweden}
\author{R.~Johansson}
\affiliation{European Spallation Source, Box 176, SE-221 00 Lund, Sweden}
\author{E.~Kasimi}
\affiliation{Department of Physics, Aristotle University of Thessaloniki, Thessaloniki, Greece}
\affiliation{Center for Interdisciplinary Research and Innovation (CIRI-AUTH), Thessaloniki, Greece}
\author{A.~Kayis Topaksu}
\affiliation{University of Cukurova, Faculty of Science and Letters, Department of Physics, 01330 Adana, Turkey}
\author{B.~Kildetoft}
\affiliation{European Spallation Source, Box 176, SE-221 00 Lund, Sweden}
\author{K.~Kordas}
\affiliation{Department of Physics, Aristotle University of Thessaloniki, Thessaloniki, Greece}
\affiliation{Center for Interdisciplinary Research and Innovation (CIRI-AUTH), Thessaloniki, Greece}
\author{A.~Leisos}
\affiliation{Physics Laboratory, School of Science and Technology, Hellenic Open University, 26335, Patras, Greece }
\author{M.~Lindroos}
\email{Deceased}
\affiliation{European Spallation Source, Box 176, SE-221 00 Lund, Sweden}
\affiliation{Department of Physics, Lund University, P.O Box 118, 221 00 Lund, Sweden}
\author{A.~Longhin}
\affiliation{Department of Physics and Astronomy "G. Galilei", University of Padova and INFN Sezione di Padova, Italy}
\author{C.~Maiano}
\affiliation{European Spallation Source, Box 176, SE-221 00 Lund, Sweden}
\author{S.~Marangoni}
\affiliation{University of Milano-Bicocca and INFN Sez. di Milano-Bicocca, 20126 Milano, Italy}
\author{C.~Marrelli}
\affiliation{CERN, 1211 Geneva 23, Switzerland}
\author{D.~Meloni}
\affiliation{Dipartimento di Matematica e Fisica, Universit\'a di Roma Tre, Via della Vasca Navale 84, 00146 Rome, Italy}
\author{M.~Mezzetto}
\affiliation{INFN Sez. di Padova, Padova, Italy}
\author{N.~Milas}
\affiliation{European Spallation Source, Box 176, SE-221 00 Lund, Sweden}
\author{J.L.~Mu\~noz}
\affiliation{Consorcio ESS-bilbao, Parque Cient\'{i}fico y Tecnol\'{o}gico de Bizkaia, Laida Bidea, Edificio 207-B, 48160 Derio, Bizkaia, Spain}
\author{K.~Niewczas}
\affiliation{Department of Physics and Astronomy, Ghent University, Proeftuinstraat 86, B-9000 Ghent, Belgium}
\author{M.~Oglakci}
\affiliation{University of Cukurova, Faculty of Science and Letters, Department of Physics, 01330 Adana, Turkey}
\author{T.~Ohlsson}
\email{Corresponding authors: S. Choubey, T. Ohlsson and S. Vihonen}
\affiliation{Department of Physics, School of Engineering Sciences, KTH Royal Institute of Technology, 106 91 Stockholm, Sweden}
\affiliation{The Oskar Klein Centre, AlbaNova University Center, Roslagstullsbacken 21, 106 91 Stockholm, Sweden}
\author{M.~Olveg{\aa}rd}
\affiliation{Department of Physics and Astronomy, FREIA Division, Uppsala University, P.O. Box 516, 751 20 Uppsala, Sweden}
\author{M.~Pari}
\affiliation{Department of Physics and Astronomy "G. Galilei", University of Padova and INFN Sezione di Padova, Italy}
\author{D.~Patrzalek}
\affiliation{European Spallation Source, Box 176, SE-221 00 Lund, Sweden}
\author{G.~Petkov}
\affiliation{Sofia University St. Kliment Ohridski, Faculty of Physics, 1164 Sofia, Bulgaria}
\author{Ch.~Petridou}
\affiliation{Department of Physics, Aristotle University of Thessaloniki, Thessaloniki, Greece}
\affiliation{Center for Interdisciplinary Research and Innovation (CIRI-AUTH), Thessaloniki, Greece}
\author{P.~Poussot}
\affiliation{IPHC, Universit\'{e} de Strasbourg, CNRS/IN2P3, Strasbourg, France}
\author{A~Psallidas}
\affiliation{Institute of Nuclear and Particle Physics, NCSR Demokritos, Neapoleos 27, 15341 Agia Paraskevi, Greece}
\author{F.~Pupilli}
\affiliation{INFN Sez. di Padova, Padova, Italy}
\author{D.~Saiang}
\affiliation{Department of Civil, Environmental and Natural Resources Engineering Lule\aa~University~of~Technology, SE-971 87 Lulea, Sweden}
\author{D.~Sampsonidis}
\affiliation{Department of Physics, Aristotle University of Thessaloniki, Thessaloniki, Greece}
\affiliation{Center for Interdisciplinary Research and Innovation (CIRI-AUTH), Thessaloniki, Greece}
\author{C.~Schwab}
\affiliation{IPHC, Universit\'{e} de Strasbourg, CNRS/IN2P3, Strasbourg, France}
\author{F.~Sordo}
\affiliation{Consorcio ESS-bilbao, Parque Cient\'{i}fico y Tecnol\'{o}gico de Bizkaia, Laida Bidea, Edificio 207-B, 48160 Derio, Bizkaia, Spain}
\author{A.~Sosa}
\affiliation{CERN, 1211 Geneva 23, Switzerland}
\author{G.~Stavropoulos}
\affiliation{Institute of Nuclear and Particle Physics, NCSR Demokritos, Neapoleos 27, 15341 Agia Paraskevi, Greece}
\author{R.~Tarkeshian}
\affiliation{European Spallation Source, Box 176, SE-221 00 Lund, Sweden}
\author{F.~Terranova}
\affiliation{University of Milano-Bicocca and INFN Sez. di Milano-Bicocca, 20126 Milano, Italy}
\author{T.~Tolba}
\affiliation{Institute for Experimental Physics, Hamburg University, 22761 Hamburg, Germany}
\author{E.~Trachanas}
\affiliation{European Spallation Source, Box 176, SE-221 00 Lund, Sweden}
\author{R.~Tsenov}
\affiliation{Sofia University St. Kliment Ohridski, Faculty of Physics, 1164 Sofia, Bulgaria}
\author{A.~Tsirigotis}
\affiliation{Physics Laboratory, School of Science and Technology, Hellenic Open University, 26335, Patras, Greece }
\author{S.~E.~Tzamarias}
\affiliation{Department of Physics, Aristotle University of Thessaloniki, Thessaloniki, Greece}
\affiliation{Center for Interdisciplinary Research and Innovation (CIRI-AUTH), Thessaloniki, Greece}
\author{G.~Vankova-Kirilova}
\affiliation{Sofia University St. Kliment Ohridski, Faculty of Physics, 1164 Sofia, Bulgaria}
\author{N.~Vassilopoulos}
\affiliation{Institute of High Energy Physics (IHEP) Dongguan Campus, Chinese Academy of Sciences (CAS), Guangdong 523803, China}
\author{S.~Vihonen}
\email{Corresponding authors: S. Choubey, T. Ohlsson and S. Vihonen}
\affiliation{Department of Physics, School of Engineering Sciences, KTH Royal Institute of Technology, 106 91 Stockholm, Sweden}
\affiliation{The Oskar Klein Centre, AlbaNova University Center, Roslagstullsbacken 21, 106 91 Stockholm, Sweden}
\author{J.~Wurtz}
\affiliation{IPHC, Universit\'{e} de Strasbourg, CNRS/IN2P3, Strasbourg, France}
\author{V.~Zeter}
\affiliation{IPHC, Universit\'{e} de Strasbourg, CNRS/IN2P3, Strasbourg, France}
\author{O.~Zormpa}
\affiliation{Institute of Nuclear and Particle Physics, NCSR Demokritos, Neapoleos 27, 15341 Agia Paraskevi, Greece}

\collaboration{ESSnuSB Collaboration}

\date{\today}

\begin{abstract}
This study provides an analysis of atmospheric neutrino oscillations at the ESSnuSB far detector facility. The prospects of the two cylindrical Water Cherenkov detectors with a total fiducial mass of 540~kt are investigated over 10 years of data taking in the standard three-flavor oscillation scenario. We present the confidence intervals for the determination of mass ordering, $\theta_{23}$ octant as well as for the precisions on $\sin^2\theta_{23}$ and $|\Delta m_{31}^2|$. It is shown that mass ordering can be resolved by $3\sigma$~CL ($5\sigma$~CL) after 4~years (10~years) regardless of the true neutrino mass ordering. Correspondingly, the wrong $\theta_{23}$ octant could be excluded by $3\sigma$~CL after 4~years (8~years) in the case where the true neutrino mass ordering is normal ordering (inverted ordering). The results presented in this work are complementary to the accelerator neutrino program in the ESSnuSB project.
\end{abstract}

\maketitle

\section{\label{sec:introduction}Introduction}

Atmospheric neutrinos are one of the most formidable neutrino sources in the Nature. Cosmic-ray interactions in the atmosphere very often result in hadronic showers that produce neutrinos as a side product. The neutrinos and antineutrinos produced in this way can have a wide range of energies and directions as they gain their origin from cosmic rays, spanning over neutrino energies from hundreds of MeV up to the PeV scale. As most of the atmospheric neutrinos traverse very long distances inside the Earth before they can be observed in any neutrino detector, atmospheric neutrinos are sensitive to effects that arise from neutrino interactions with matter.

The standard theory of three-flavour neutrino oscillations states that the mixing of three active neutrinos can be parameterized with three mixing angles $\theta_{12}$, $\theta_{13}$ and $\theta_{23}$, two mass-squared differences $\Delta m_{21}^2 \equiv m_2^2 - m_1^2$ and $\Delta m_{31}^2 \equiv m_3^2 - m_1^2$, and one charge-parity (\emph{CP}) phase $\delta_{CP}$. Experimental efforts to study neutrino oscillations with neutrinos of accelerator, reactor, atmospheric and solar origin have determined the values of the mixing angles and the mass-squared differences within 1\%--5\% precision at $1\sigma$ confidence level (CL) and hinted that $\delta_{CP}$ may be $CP$-violating~\cite{Esteban:2020cvm}. Next-generation neutrino oscillation experiments aim to determine whether the neutrino masses $m_1$, $m_2$ and $m_3$ follow the normal ordering, $\Delta m_{31}^2 > 0$, or the inverted ordering, $\Delta m_{31}^2 < 0$, by studying neutrino oscillations with both neutrinos and antineutrinos. It is also to be discovered whether the \emph{CP} phase is \emph{CP}-violating, $\sin \delta_{CP} \neq 0$, or \emph{CP}-conserving, $\sin \delta_{CP} = 0$, and whether the mixing angle $\theta_{23}$ resides in the low octant, $\theta_{23} < 45^\circ$, or the high octant, $\theta_{23} > 45^\circ$. Future neutrino oscillation experiments will also test the precision of the Standard Model by looking for non-standard interactions and additional neutrino families.

The European Spallation Source neutrino SuperBeam (ESSnuSB) project~\cite{Alekou:2022emd} aims to study leptonic $CP$ violation by sending high-power neutrino and antineutrino beams over a baseline length that is 360~km long. The main source of neutrinos in this project would be the ESS linear accelerator, which is capable of creating ultra-pure muon neutrino beams with 5~MW output. The advantage of ESSnuSB would be its access to neutrino oscillations at the second oscillation maximum, which is expected to have significant potential to accurately measure the value of $\delta_{CP}$. The second oscillation maximum is also expected to enable measurements on the standard neutrino oscillation parameters with high precision~\cite{ESSnuSB:2023ogw}. The prospects of ESSnuSB also include other physics cases related to neutrinos, such as the coherent elastic neutrino-nucleus scattering~\cite{Baxter:2019mcx,Chatterjee:2022mmu} and searches for physics beyond the Standard Model~\cite{Blennow:2015nxa,Ghosh:2019zvl,Blennow:2020snb,Choubey:2020dhw,Majhi:2021api,Chatterjee:2021xyu,Cheng:2022lys,Cordero:2022fwb,ESSnuSB:2023lbg,ESSnuSB:2024yji}.

In the present work, we examine the prospects of measuring atmospheric neutrino oscillations at the ESSnuSB far detector facility. The ESSnuSB far detector utilizes the Water Cherenkov technology, where neutrino properties are reconstructed by observing the Cherenkov light that is emitted from neutrino interactions with water. The far detector facility is planned to consist of two identical water cylinders that would be placed inside the mine in Zinkgruvan in central Sweden at the depth of 1~km. The combined fiducial mass of the proposed far detector facility is $540$~kt, which would make the ESSnuSB far detectors approximately $2.9$ times larger than the Hyper-Kamiokande detector~\cite{Hyper-Kamiokande:2018ofw}. The geographical location of Zinkgruvan has a relatively high flux of atmospheric neutrinos thanks to its proximity to the North Pole, making the conditions at ESSnuSB far detectors promising for the study of atmospheric neutrino oscillations. The experimental program of ESSnuSB would complement the prospects of the currently planned neutrino experiments such as DUNE~\cite{DUNE:2015lol}, Hyper-Kamiokande~\cite{Hyper-Kamiokande:2018ofw}, IceCube-Gen2~\cite{IceCube-Gen2:2020qha} and KM3NeT~\cite{KM3Net:2016zxf}.

The numerical study of atmospheric neutrino oscillations is carried out as follows. We first generate a large set of Monte Carlo (MC) samples for atmospheric neutrino interactions at ESSnuSB far detectors using the neutrino event generator GENIE~\cite{Andreopoulos:2009rq,GENIE:2021npt}. An in-house written analysis software based on Python is then used to emulate detector response in the ESSnuSB far detectors and compute sensitivities to neutrino mass ordering, $\theta_{23}$ octancy and precisions on $\sin^2\theta_{23}$ and $\Delta m_{31}^2$ assuming a $5.4$~Mt$\cdot$year total exposure. The neutrino oscillation probabilities used in this analysis are calculated numerically with the General Long-Baseline Experiment Simulator (GLoBES)~\cite{Huber:2004ka,Huber:2007ji}. The results obtained in this work are complementary to the accelerator physics program of the ESSnuSB project\footnote{The complementarity between atmospheric and accelerator neutrinos at ESSnuSB was previously studied in Ref.~\cite{Blennow:2019bvl}, where the authors assumed a MEMPHYS-like detector with 1~Mt fiducial mass and the atmospheric neutrino fluxes of Gran Sasso.}.

This article is divided into the following sections. Section~\ref{sec:oscillations} provides a brief review of atmospheric neutrino oscillations in vacuum and matter. The ESSnuSB far detector complex as well the atmospheric neutrino flux are described in section~\ref{sec:configuration}. Section~\ref{sec:analysis} presents the analysis techniques used on the MC samples, while the major numerical results are shown in section~\ref{sec:results}. We finally provide concluding remarks in section~\ref{sec:summary}.

\section{\label{sec:oscillations}Neutrino oscillations}

The concept of neutrino oscillations is a quantum phenomenon that derives from the non-conventional nature of neutrino mass. It is known that neutrino mass states and flavour states do not coincide, leading to the possibility that a neutrino born in flavour state $\nu_\alpha$ may be found in a different flavour state $\nu_\beta$ after propagating distance $L$ with energy $E_\nu$. This mixing between neutrino flavour and mass eigenstates can be described with a complex unitary matrix,
\begin{equation}
    \label{eq:2.1}
    |\nu_\alpha\rangle = \sum_{i=1}^{3} U_{\alpha i}^{*} |\nu_i\rangle,
\end{equation}
where $\alpha = e$, $\mu$ or $\tau$. Here $|\nu_i\rangle$ are eigenstates in the mass basis and $|\nu_\alpha\rangle$ are the states in the flavour basis, respectively. In the standard parametrization of leptonic mixing, the mixing matrix $U$ is the so-called Pontecorvo-Maki-Nakagawa-Sakata (PMNS) matrix and it is given by
\begin{equation}
    \label{eq:2.2}
    U =
    \begin{pmatrix}
        1 & 0 & 0\\
        0 & c_{23} & s_{23}\\
        0 & -s_{23} & c_{23}
    \end{pmatrix}
    \begin{pmatrix}
        c_{13} & 0 & s_{13} e^{-i\delta_{CP}}\\
        0 & 1 & 0\\
        -s_{13} e^{-i\delta_{CP}} & 0 & c_{13}
    \end{pmatrix}    
    \begin{pmatrix}
        c_{12} & s_{12} & 0\\
        -s_{12} & c_{12} & 0\\
        0 & 0 & 1
    \end{pmatrix},
\end{equation}
where $c_{ij}$ and $s_{ij}$ are defined as $\cos \theta_{ij}$ and $\sin \theta_{ij}$, respectively. The parametrization in the PMNS matrix~(\ref{eq:2.2}) involves three leptonic mixing angles $\theta_{12}$, $\theta_{13}$ and $\theta_{23}$ and one \emph{CP} phase $\delta_{CP}$. When neutrinos propagate in vacuum, the neutrino oscillation probabilities can be computed from the time-evolution operator as $P_{\nu_\alpha \rightarrow \nu_\beta} (E_\nu, L) = |S|^2 \equiv |e^{-i H_0 L}|^2$, where $S \equiv e^{-i H_0 L}$ is the evolutionary operator and the vacuum Hamiltonian $H_0$ is given by
\begin{equation}
    \label{eq:2.3a}
    H_0 = \frac{1}{2E_\nu}U
    \begin{pmatrix}
        0 & 0 & 0\\
        0 & \Delta m_{21}^2 & 0\\
        0 & 0 & \Delta m_{31}^2\\
    \end{pmatrix}
    U^\dagger,
\end{equation}
and the full probability formula can be written as
\begin{equation}
    \label{eq:2.3}
    P_{\nu_\alpha  \rightarrow \nu_\beta} (E_\nu, L) = \delta_{\alpha \beta} - 4 \sum_{i>j}\mathcal{R}\left(U_{\alpha i}^* U_{\beta i} U_{\alpha j} U_{\beta j}^*\right) \sin^2\Delta_{ij} \pm \sum_{i>j}\mathcal{I}\left(U_{\alpha i}^* U_{\beta i} U_{\alpha j} U_{\beta j}^*\right)\sin2\Delta_{ij}.
\end{equation}
The quantity $\Delta_{ij} \equiv L \Delta m_{ij}^2/(4E_\nu)$ in equation~(\ref{eq:2.3}) defines the oscillation mode and the sign of the second term is positive for neutrinos and negative for antineutrinos. Neutrino oscillations in vacuum therefore depend on six independent parameters, now including the two mass-squared differences $\Delta m_{21}^2$ and $\Delta m_{31}^2$ in addition to the mixing angles and the \emph{CP} phase.

In atmospheric neutrino oscillations, the relevant oscillation channels are $\nu_e \rightarrow \nu_e$, $\nu_\mu \rightarrow \nu_\mu$, $\nu_e \rightarrow \nu_\mu$ and $\nu_\mu \rightarrow \nu_e$. When $L / E_\nu \ll$ 1, the oscillation probabilities are driven by the $\Delta_{31}$ mode, since $\Delta_{21} \ll$ 1. In this case, the neutrino oscillation probabilities can be approximated with the analytical formulas~\cite{Akhmedov:2004ny,Super-Kamiokande:2017yvm}
\begin{align}
    P_{\nu_e \rightarrow \nu_e} (E_\nu, L) & \simeq 1 - \sin^2 2\theta_{13} \sin^2 \left( \frac{L\Delta m_{31}^2}{4\,E_\nu} \right),\label{eq:2.4a}\\
    P_{\nu_\mu \rightarrow \nu_\mu} (E_\nu, L) & \simeq 1 - 4\cos^2\theta_{13}\sin^2\theta_{23} (1 - \cos^2\theta_{13}\sin^2\theta_{23}) \sin^2 \left( \frac{L \Delta m_{31}^2}{4\,E_\nu} \right),\label{eq:2.4b}\\
    P_{\nu_\mu \leftrightarrow \nu_e} (E_\nu, L) & \simeq \sin^2 \theta_{23} \sin^2 2\theta_{13} \sin^2 \left( \frac{L \Delta m_{31}^2}{4\,E_\nu} \right)\label{eq:2.4c},
\end{align}
where the neutrino oscillation probabilities are given at zeroth-order in the small parameter $\Delta m_{21}^2 / \Delta m_{31}^2$. Equation~(\ref{eq:2.4c}) gives the parameter dependence of the neutrino oscillation probabilities for both the $\nu_\mu \rightarrow \nu_e$ and $\nu_e \rightarrow \nu_\mu$ channels. As expected, the oscillations between muon and electron neutrino states are driven by the leptonic mixing parameters $\theta_{23}$ and $\Delta m_{31}^2$, and also by the mixing angle $\theta_{13}$. On the other hand, the sensitivity to the $CP$ phase $\delta_{CP}$ arises from the sub-leading terms that are not present in formulas~(\ref{eq:2.4a})--(\ref{eq:2.4c}).

The majority of atmospheric neutrinos are created about 15~km above the Earth's surface. Atmospheric neutrinos may therefore undergo distances between 15~km and 12~742~km, the latter of which is equivalent to neutrinos passing through the full diameter of the Earth. Taking the matter effects into account, the effective Hamiltonian can be written in the mass basis as
\begin{equation}
    \label{eq:2.5}
    H_m = \frac{1}{2E_\nu}
    \begin{pmatrix}
        m_1^2 & 0 & 0\\
        0 & m_2^2 & 0\\
        0 & 0 & m_3^2\\
    \end{pmatrix}
    + U^\dagger
    \begin{pmatrix}
        a & 0 & 0\\
        0 & 0 & 0\\
        0 & 0 & 0\\
    \end{pmatrix}
    U,
\end{equation}
where $a = \pm \sqrt{2} G_F N_e$ is positive for neutrinos and negative for antineutrinos, $G_F$ is the Fermi coupling constant and $N_e$ is the number density of electrons in the Earth. The mixing matrix $U$ is the PMNS matrix we defined in equation~(\ref{eq:2.2}). For constant matter density, the neutrino interactions with matter can be described with effective mixing parameters which reduce the oscillation probabilities back to the vacuum formulas we provided in equations~(\ref{eq:2.4a})--(\ref{eq:2.4c}). This is achieved through the following transformations:
\begin{align}
    \Delta m_{31}^2 & \rightarrow \Delta m_{31}^2 \sqrt{\sin^2 2\theta_{13} + (\Gamma - \cos 2\theta_{13})^2}\label{eq:2.6a},\\
    \sin^2 2\theta_{13} & \rightarrow \frac{\sin^2 2\theta_{13}}{\sin^2 2\theta_{13} + (\Gamma - \cos 2\theta_{13})^2}\label{eq:2.6b},
\end{align}
where we define $\Gamma = a E_\nu / \Delta m_{31}^2$. One can readily see from equations~(\ref{eq:2.6a}) and~(\ref{eq:2.6b}) that the effective mixing is maximal when $\Gamma = \cos 2\theta_{13}$ and the neutrino oscillation probabilities are significantly enhanced. Neutrinos undergoing these conditions are therefore said to go though resonant transition.

As the distances that atmospheric neutrinos can traverse inside the Earth vary significantly, the constant matter density approach is not applicable and a detailed matter density profile must be used. The effective operator describing the propagation of neutrino mass eigenstates in $N$ layers of constant matter density can be written as~\cite{Barger:1980tf,Langacker:1982ih}
\begin{equation}
    \label{eq:06}
    X = \sum_k \left[ \prod_{j \neq k} \frac{2 E_\nu H_{m} - m_j^2 I}{m_k^2 - m_j^2} \right] e^{-i m_k^2 L / (2 E_\nu)},
\end{equation}
where $k = 1, 2, \ldots, N$ and $I$ is the identity matrix. Equation~(\ref{eq:06}) presents the propagated eigenvalues $m_i^2 / (2E_\nu)$ of the constant matter density Hamiltonian $H_{m}$. The rows of the matrix $X$ then represent the propagated eigenvectors of the neutrino mass matrix. The neutrino oscillation probabilities that take into account neutrino interactions with matter can therefore be obtained from the formula
\begin{equation}
    \label{eq:2.7}
    P_{\nu_\alpha \rightarrow \nu_\beta}(E_\nu, L) = \left| (U X U^\dagger)\right|^2,
\end{equation}
where $U$ is the PMNS matrix defined in equation~(\ref{eq:2.2}) and $\alpha, \beta = e, \mu$ and $\tau$. A convenient way to approximate the varying matter density is to implement the Preliminary Reference Earth Model (PREM)~\cite{Dziewonski:1981xy}, which treats the internal structure of the Earth as a finite number of layers with constant matter density. To calculate the matter density for a given propagation distance inside the Earth, one must determine the zenith angle $\theta_{z}$ of the incoming neutrino. The full three-flavour neutrino oscillation probabilities can then be calculated as
\begin{equation}
    \label{eq:2.8}
    P_{\nu_\alpha \rightarrow \nu_\beta}(E_\nu, h, \theta_z) = \left| \left(U \prod_{i}^{N} X(L_i, \rho_i, E_\nu) U^\dagger \right)_{\alpha\beta}\right|^2,
\end{equation}
where $h$ is the production height at which the atmospheric neutrino is produced. Note that the neutrino trajectory is now sliced into $N$ layers such that $L = \sum_i L_i$. Each layer is assumed to have a constant matter density $\rho_i$.

\section{\label{sec:configuration}The ESSnuSB far detectors}

The ESSnuSB far detector facility is planned to consist of two identical Water Cherenkov detectors. The far detectors have the shape of standing cylinders with a height of 76~m and a diameter of 76~m. Each cylinder is expected to hold ultra-pure water of about 270~kt fiducial mass, giving the total fiducial mass of the far detector complex as 540~kt. The cylinders are to be instrumented with photo-multiplier tubes (PMTs), which would reduce the fiducial volume by 2~m from the cylindrical surface. The PMT structure is planned to feature inward-pointing 20-inch PMTs with the purpose of detecting Cherenkov light from charged particles that are produced in neutrino interactions. There would also be outward-pointing PMTs to be used as a veto. The area covered by the inward-pointing PMTs would form 30\% coverage. After the fiducial cuts, the active volume of the detectors has the dimensions of 70~m height and 70~m diameter. 

The chosen site for the ESSnuSB far detector complex is the mine in Zinkgruvan in central Sweden. The 1100~m rock overburden provides a sufficient protection from the main backgrounds to the experiment's accelerator neutrino program. The mine also shields the detectors from the cosmic muons and muons that are created by beam neutrino interactions in the rock. More information on the backgrounds can be found in Ref.~\cite{Alekou:2022emd}. The atmospheric neutrino flux at this location is expected to be similar to that of the Pyh\"asalmi mine in Finland and several percent higher than the atmospheric neutrino flux in Kamioka in Japan~\cite{Gaisser:2002jj,Honda:2015fha}. The atmospheric neutrino flux is the highest in the tangential plane on the Earth's surface, $\cos \theta_z = 0$, and decreases towards the directions where the neutrino flux is perpendicular, $\cos \theta_z = \pm 1$. The muon-to-electron flavour ratio of the atmospheric neutrinos is about 2:1, with a large number of the neutrinos carrying sub-GeV energies. We compute the atmospheric neutrino fluxes from the average of the respective fluxes that correspond to the solar minimum and the solar maximum periods\footnote{The flux files are available in \texttt{http://www-rccn.icrr.u-tokyo.ac.jp/mhonda/public/}.}.

\begin{figure}[!t]
        \center{\includegraphics[width=0.6\textwidth]{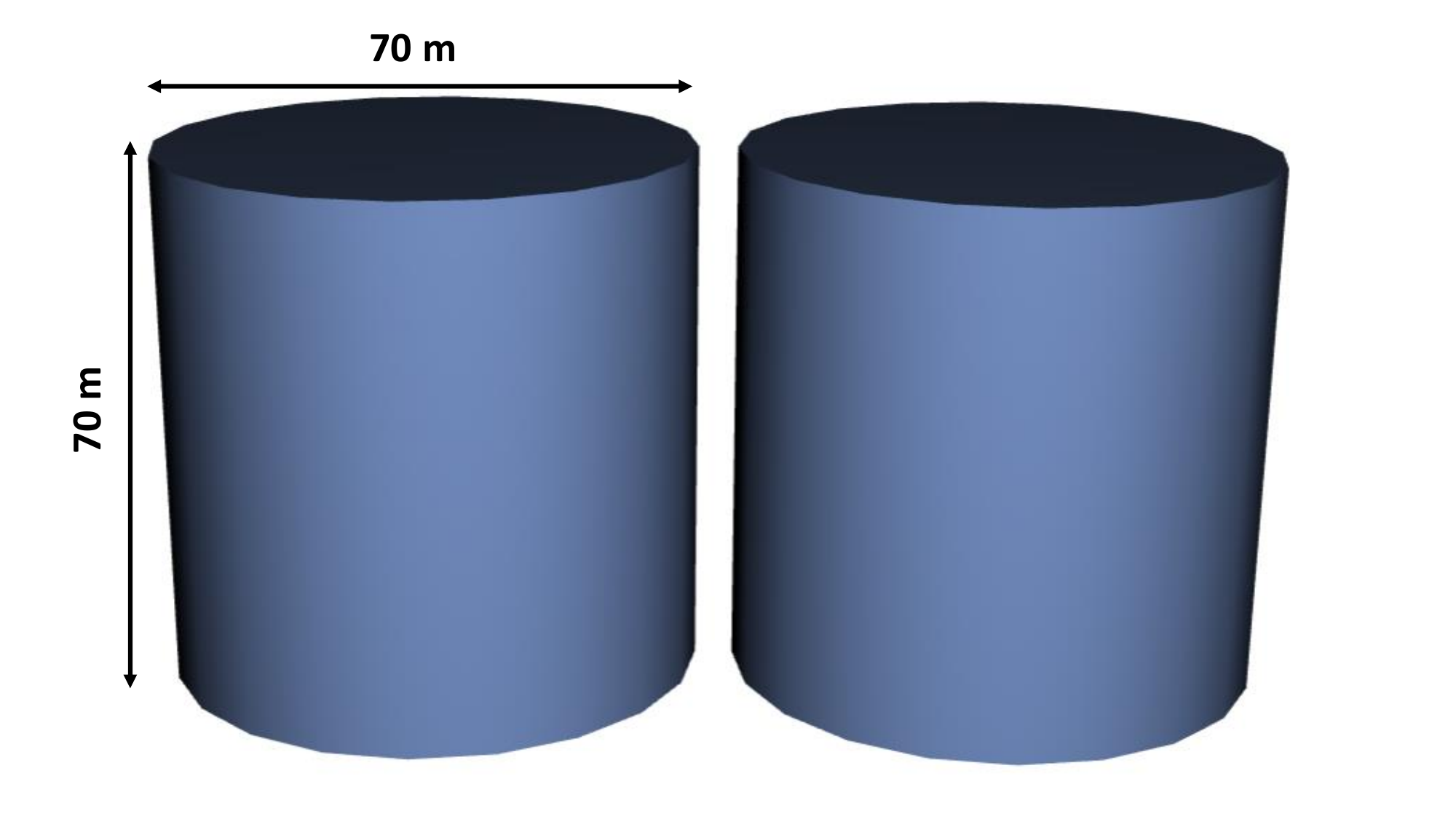}}
        \caption{Illustration of the detector fiducial volume geometry used to describe the far detector setup in ESSnuSB. The detector active target consists entirely of ultra-pure water with 88.89\% and 11.11\%  of $^{16}$O and $^1$H atoms, respectively. The total fiducial mass of the two-detector complex is 540~kt. The image was created using the ROOT geometry package~\cite{Brun:1997pa}.}
        \label{fig:geometry}
\end{figure}

In this work, the performance of the ESSnuSB far detector complex is studied in the context of atmospheric neutrino detection. The neutrino events emerging from atmospheric neutrino interactions are generated with the GENIE~\cite{Andreopoulos:2009rq} event generator, whereas the detector geometry is created using the ROOT geometry package~\cite{Brun:1997pa}. An illustration of the detector geometry can be found in Figure~\ref{fig:geometry}. The total fiducial mass of the far detector complex is taken to be 540~kt, which can be achieved by creating a ROOT geometry of two cylindrical volumes with 70~m diameter and 70~m height each. The detector material is taken to be ultra-pure water, where 88.89\% of active mass is formed by $^{16}$O atoms and 11.11\% by $^1$H atoms. The PMT structure and the cylindrical containers are not taken into account in the MC event generation and are therefore neglected in the ROOT geometry.
The main backgrounds to the atmospheric neutrinos include the cosmic muons as well as the muons that are created by neutrino interactions inside the rock. Beam neutrinos can also be treated as background to atmospheric neutrinos. The aforementioned backgrounds are estimated to be small and are not taken into account in our analysis.

\section{\label{sec:analysis}Atmospheric neutrino analysis}

The statistical analysis of the simulated neutrino events is given as follows. The GENIE event generator~\cite{Andreopoulos:2009rq} is used to generate MC samples which are processed to obtain neutrino events for the test and the true hypotheses. The processed MC events are then analysed with the likelihood function:
\begin{equation}
    \label{eq:4.1}
    \chi^2 = 2 \sum_{n=1}^{2000} \left( E_n - O_n + O_n \log \frac{O_n}{E_n} \right) + \sum_{i=1}^{5} \left( \frac{\zeta_i}{\sigma_i} \right)^2.
\end{equation}
Here $E_n$ and $O_n$ correspond to the expected and observed neutrino events in the $n^{\rm th}$ analysis bin and $\zeta_i$ is the nuisance parameter modeling the $i^{th}$ systematic uncertainty with standard deviation $\sigma_i$. The systematic uncertainties are treated with the pull method~\cite{Fogli:2002pt}, where
\begin{equation}
    \label{eq:4.2}
    E_n = E_{n, 0} \left( 1 + \sum_{i=1}^{5} f_{i, n} \zeta_i \right).
\end{equation}
In this equation, $E_{n, 0}$ represents the original MC expectation and $f_{i, n}$ is the coefficient that determines the weight of the pull parameter $\zeta_i$. The analysis bins $n$ runs through 100 energy and 20 cosine zenith angle bins, which are distributed evenly over neutrino energies $E_\nu\in[0.1,100]$~GeV and neutrino cosine zenith angles $\cos \theta_{z}\in[-1,1]$. In order to assess the effect of detector response, the MC events undergo a folding process, where Gaussian smearing is applied to the neutrino energies $E_\nu$ and also to the neutrino cosine zenith angles $\cos \theta_{z}$ in each analysis bin. It is assumed that the migration between the $E_\nu$ bins and $\cos \theta_{z}$ bins is small. The Gaussian smearing is therefore carried out independently for the $E_\nu$ bins and the $\cos \theta_z$ bins. The number of atmospheric neutrino events in each analysis bin is furthermore multiplied by the appropriate detector efficiency. Similar techniques have been used for Water Cherenkov detectors in {\em e.g.} Ref.~\cite{Kelly:2017kch}. As the Water Cherenkov technology planned for the ESSnuSB far detectors is not sensitive to the sign of the electric charge of primary leptons, the neutrino and antineutrino events of the same lepton flavor are analyzed without distinction between $CP$ charges\footnote{Some sensitivity to the $CP$ charge of neutrinos and antineutrinos could be accomplished by via gadolinium-doping, which has been successfully implemented in Super-Kamiokande, see Ref.~\cite{Super-Kamiokande:2021the}. Investigations are currently underway to include gadolinium-doping in ESSnuSB far detectors, which could benefit from the treatment.}. We adopt the detector efficiencies from Ref.~\cite{Alekou:2022emd}, taking into account leptonic flavor and $CP$ charge of each atmospheric neutrino. We furthermore use 30\% resolution for sub-GeV neutrino energies and 10\% resolution for multi-GeV neutrino energies, respectively, and a constant 10$^\circ$ resolution for the neutrino cosine zenith angles. We have checked that these resolution functions reproduce rather well the mass ordering results obtained by the Hyper-Kamiokande collaboration with the atmospheric neutrino fluxes at Kamioka in Japan and the detector size normalized to the Hyper-Kamiokande detector~\cite{Hyper-Kamiokande:2018ofw}.

The neutrino oscillations are taken into account by applying the so-called reweighing method. For each MC event, the oscillation probabilities are computed according to the initial energy and cosine zenith angle of the associated neutrino. The reweighing is carried out by assigning each MC event a random number $S \in$ [0, 1], which is then compared to the relevant neutrino oscillation probability. For instance, if a muon neutrino event is assigned a random number $S$ that satisfies $S < P_{\nu_\mu \rightarrow \nu_e}$, the event is classified as an oscillated electron neutrino event. Correspondingly, the event is classified as an unoscillated muon neutrino event if $P_{\nu_\mu \rightarrow \nu_e} < S < P_{\nu_\mu \rightarrow \nu_e} + P_{\nu_\mu \rightarrow \nu_\mu}$ and an oscillated tau neutrino event if $S > P_{\nu_\mu \rightarrow \nu_e} + P_{\nu_\mu \rightarrow \nu_\mu}$, respectively. Electron neutrino events as well as electron antineutrino and muon antineutrino events are treated analogously. The matter densities are interpolated from PREM, which is evaluated with 81 layers to ensure sufficient detail in the matter effects. We resolve the neutrino propagation distances from their corresponding zenith angles with the following equation
\begin{equation}
    \label{eq:4.3}
    L = \sqrt{(R + h)^2 - (R - d)^2 \sin^2 \theta_z} - (R - d)\cos \theta_z,
\end{equation}
where $R$ is the radius of the Earth, $h$ is the height where atmospheric neutrinos are born and $d$ is the depth of the neutrino detector. In our analysis, we assume the detector to be located in the mine at $d =$ 1~km. We furthermore assume the neutrino production height $h$ to be 15~km and the diameter of the Earth $R$ to be 6371~km, respectively. The probability calculation is executed with GLoBES, while the reweighing method is implemented in the main analysis code.

\begin{table}[!t]
\begin{center}
\begin{tabular}{|c|c|}\hline
{\bf Systematic error} & {\bf Uncertainty} \\ \hline
\rule{0pt}{3ex}Flux normalization & 20\% \\ 
\rule{0pt}{3ex}Cross-section normalization & 10\% \\ 
\rule{0pt}{3ex}Zenith angle dependence & varies \\ 
\rule{0pt}{3ex}Energy tilt & varies \\ 
\rule{0pt}{3ex}Detector & 5\% \\ \hline
\end{tabular}
\end{center}
\caption{\label{tab:systematics}List of systematic uncertainties used in this work. The methodology is adopted from Refs.~\cite{Ghosh:2012px}. See the text for the implementation of the zenith angle dependence and energy tilt errors.}
\end{table}

The evaluation of the systematic uncertainties is based on the approach described in Ref.~\cite{Ghosh:2012px}. The systematic uncertainties are evaluated with the pull parameters $\zeta_1, \zeta_2, \zeta_3, \zeta_4$ and $\zeta_5$ in equation~(\ref{eq:4.2}). Each analysis bin is assigned weights $f_{1, n}, f_{2, n}, f_{3, n}, f_{4, n}$ and $ f_{5, n}$ that determine the impact from individual pull parameters in the $n^{th}$ analysis bin.  In the analysis of the MC events generated for ESSnuSB far detectors, there are five different uncertainties that influence the analysis of atmospheric neutrinos in ESSnuSB: (i) flux normalization error, (ii) cross-section normalization error, (iii) zenith angle dependence error, (iv) energy tilt error and (v) detector error. The summary of the systematic uncertainties and the values used in this work is given in Table~\ref{tab:systematics}. Systematic errors (i) and (ii) are ordinary normalization errors derived from uncertainty in the atmospheric neutrino fluxes and cross-sections, respectively. The zenith angle uncertainty (iii) arises from the uncertainty on the zenith angle bins and it depends on the value of neutrino $\cos \theta_z$. The energy tilt error (iv) is calculated directly from the ratio of atmospheric neutrino fluxes. The detector uncertainty (v) is considered to be a normalization error. These systematic uncertainties are treated as uncorrelated errors.

We now discuss the systematic uncertainties related to zenith angle dependence and energy tilt errors. The zenith angle error arises from the uncertainty in the zenith angle binning. It is independent of the neutrino energy and it can be calculated directly from the value of the neutrino cosine zenith angle $\cos \theta_z$. In this work, we calculate the weights for zenith angle dependence as 5\% of the neutrino $\cos \theta_z$ value. The error weights associated with the zenith angle dependence therefore belong to the interval $f_{3, n} \in [-5\%, 5\%]$. On the other hand, the energy tilt error takes into account potential deviations from the power law dependence of the atmospheric neutrino fluxes. We treat the energy tilt error using the method discussed in Ref.~\cite{Gonzalez-Garcia:2004pka}. In this approach, the MC events are generated using atmospheric neutrino fluxes that have been perturbed by a small deviation $\delta$ (called the tilt error) from the standard atmospheric neutrino flux spectrum. The distorted fluxes $\Phi_\delta$ are computed from the original fluxes $\Phi_0$ as
\begin{equation}
    \label{eq:4.4}
    \Phi_\delta (E) = \Phi_0 (E) \left( \frac{E}{E_{0}} \right)^\delta \simeq \Phi_0 (E) \left( 1 + \delta \log\frac{E}{E_0} \right),
\end{equation}
where $E$ is the neutrino energy in the distorted spectrum and $E_0$ is a reference upon which the power-law deviation is imposed. The error weights $f_{4, n}$ are then extracted for every analysis bin $n$ by comparing the generated MC samples at the original and distorted scales. Following the example in Ref.~\cite{Gonzalez-Garcia:2004pka}, we obtained the distorted fluxes at $\delta =$ 5\% and reference energy $E_0 =$ 2~GeV. The weights were determined for the tilt error by generating MC samples for 100 years of ESSnuSB far detector exposure using both the nominal and the distorted atmospheric neutrino fluxes and calculating the relative difference in the MC samples. We estimate the resulting weights to fall mostly within the range $f_{4, n} \in [-5\%, 5\%]$ for all analysis bins.

We perform the analysis of the generated MC samples with a grid scan. The parameters $\theta_{12}$ and $\Delta m_{21}^2$ are fixed at $\sin^2 \theta_{12} =$ 0.303, $\Delta m_{21}^2 =$ 7.41$\times$10$^{-5}$~eV$^2$ in the scan, whereas the parameters $\theta_{23}$ and $\Delta m_{31}^2$ are varied. The values are adopted from \texttt{NuFit 5.2}~\cite{Esteban:2020cvm,NuFIT:5.2} assuming normal ordering while including the atmospheric neutrino data from Super-Kamiokande. The mixing angle $\theta_{13}$ is fixed according to the reactor neutrino measurements at $\sin^2\theta_{13} = 0.02225$. We additionally scan the \emph{CP} phase $\delta_{CP}$ over range $[0, 2\pi)$. The scan ranges are listed in Table~\ref{tab:scanrange}. In some instances, we present our results as functions of $\sin^2\theta_{23}$ values that are used in the true hypothesis. In those cases, the true values of $\sin^2\theta_{23}$ are reported separately.
\begin{table}[!t]
\begin{center}
\begin{tabular}{|c|c|c|c|}\hline
{\bf Scan parameter} & {\bf True value} & {\bf Scan range} & {\bf Scan points} \\ \hline
\rule{0pt}{3ex}$\sin^2 \theta_{12}$ & $0.303$ & $0.303$ & fixed \\ 
\rule{0pt}{3ex}$\sin^2 \theta_{13}$ & $0.02225$ & $0.02225$ & fixed \\ 
\rule{0pt}{3ex}$\sin^2 \theta_{23}$ & $0.451$ & $$[0.4, 0.6]$$ & 50 points \\ 
\rule{0pt}{3ex}$\delta_{CP}$ & $1.29\pi$ & $[0, 2\pi)$ & 4 points \\ 
\rule{0pt}{3ex}$\Delta m_{21}^2$ & $7.41\times$10$^{-5}$~eV$^2$ & $7.41\times$10$^{-5}$~eV$^2$ & fixed \\ 
\rule{0pt}{3ex}$|\Delta m_{31}^2|$ & $2.507\times10^{-3}$eV$^2$ & $[2.40, 2.60]\times$10$^{-3}$~eV$^2$ & 50 points \\ \hline
\end{tabular}
\end{center}
\caption{\label{tab:scanrange}Values of the neutrino oscillation parameters used in this work. The true values are adopted from \texttt{NuFit 5.2} assuming normal ordering with the atmospheric neutrino data from Super-Kamiokande~\cite{Esteban:2020cvm, NuFIT:5.2} included.} The scan ranges are also shown for the neutrino oscillation parameters.
\end{table}

\section{\label{sec:results}Numerical results}

\begin{figure}[!t]
        \center{\includegraphics[width=0.90\textwidth]{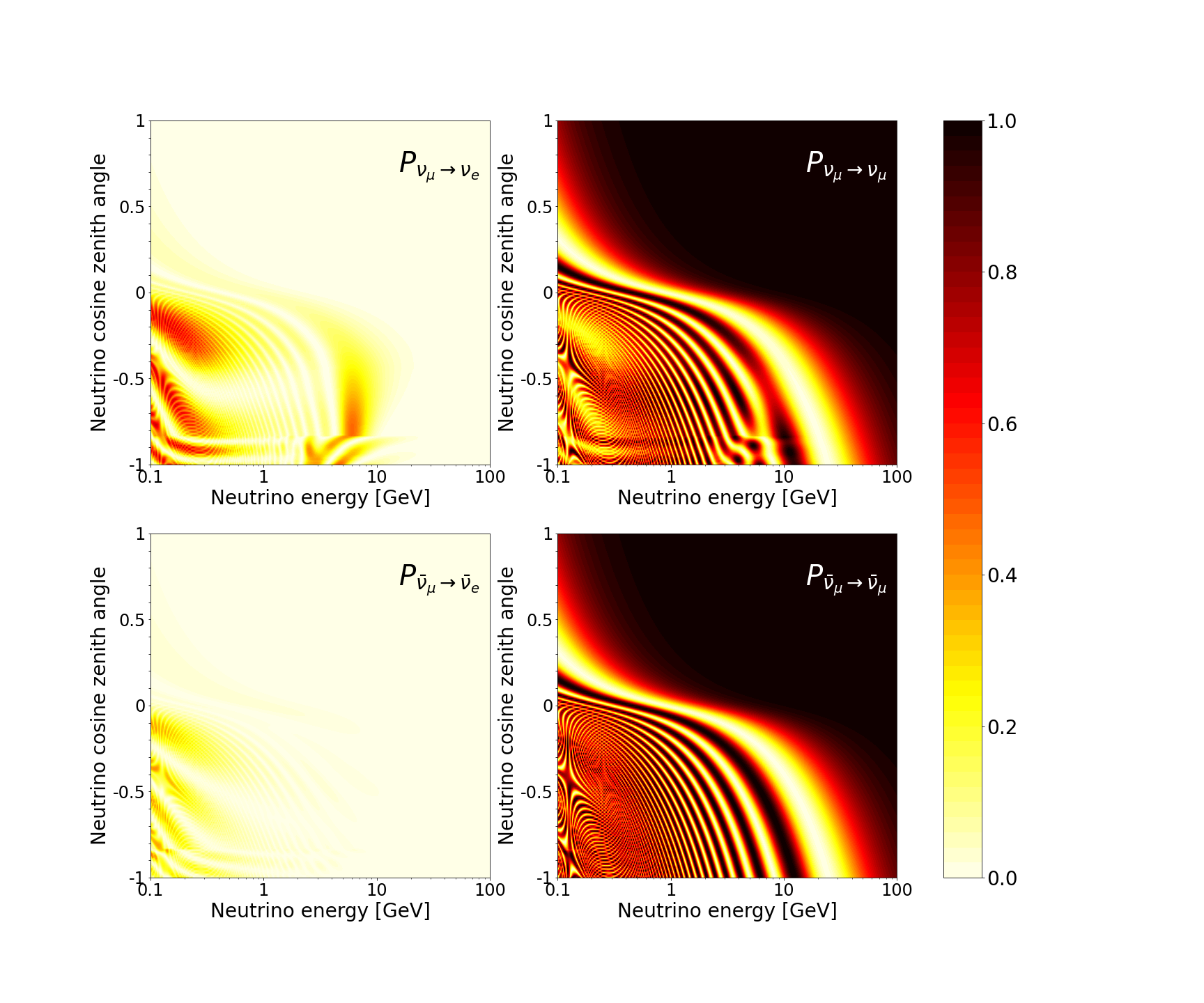}}
        \vspace{-1.0cm}
        \caption{Neutrino oscillation probabilities for $\nu_\mu \rightarrow \nu_e$ (top-left), $\nu_\mu \rightarrow \nu_\mu$ (top-right), $\bar{\nu}_\mu \rightarrow \bar{\nu}_e$ (bottom-left) and $\bar{\nu}_\mu \rightarrow \bar{\nu}_\mu$ (bottom-right) channels as function of neutrino energy and neutrino cosine zenith angle. The probabilities were calculated at the global best-fit values of the neutrino oscillation parameters assuming normal ordering~\cite{Esteban:2020cvm}.}
        \label{fig:oscillograms}
\end{figure}

The analysis of the MC samples is carried out with pre-computed neutrino oscillation probabilities. Figure~\ref{fig:oscillograms} presents neutrino oscillation probabilities in the oscillation channels $\nu_\mu \rightarrow \nu_e$, $\nu_\mu \rightarrow \nu_\mu$, $\bar{\nu}_\mu \rightarrow \bar{\nu}_e$ and $\bar{\nu}_\mu \rightarrow \bar{\nu}_\mu$. The neutrino oscillation probabilities are provided as functions of neutrino energy $E_\nu$ and neutrino cosine zenith angle $\cos \theta_z$. The probabilities were calculated using the neutrino oscillation parameter values that are given in Table~\ref{tab:scanrange}. The matter effects were included in the probability calculation by using 81 constant matter density layers which have been obtained from PREM. The color coding in Figure~\ref{fig:oscillograms} indicates the values of neutrino oscillation probabilities, where the darker shades represent high neutrino oscillation probabilities and lighter areas low probabilities. The differences between the neutrino and antineutrino oscillation probabilities can be observed in the probabilities computed for the conversion channels $\nu_\mu \rightarrow \nu_e$ and $\bar{\nu}_\mu \rightarrow \bar{\nu}_e$, which are presented in the top-left and the bottom-left panels, respectively. The most interesting features in these two panels are found in the segment $\cos \theta_z \in [-1, -0.8]$, which corresponds to atmospheric neutrinos that traverse near the full diameter of the Earth. The neutrino oscillation probabilities belonging to this segment reach the first local maximum for $\nu_\mu \rightarrow \nu_e$ at the neutrino energies $E_\nu \sim 2$~GeV, while the antineutrino channel $\bar{\nu}_\mu \rightarrow \bar{\nu}_e$ shows no significant neutrino oscillation probabilities in the same region. This is an artifact of matter effects enhancing oscillations in the neutrino channel for the normal neutrino mass ordering. Moreover, neutrino energies $E_\nu \sim 6$~GeV display notable distortions in the neutrino disappearance channel $\nu_\mu \rightarrow \nu_\mu$ but not in the antineutrino disappearance channel $\bar{\nu}_\mu \rightarrow \bar{\nu}_\mu$. This distortion is due to the matter effects in the neutrino channel for the normal neutrino mass ordering. Changing the sign of $\Delta m_{31}^2$ would switch the role of matter effects in the oscillation probabilities in the neutrino channels $\nu_\mu \rightarrow \nu_e$ and $\nu_\mu \rightarrow \nu_\mu$ and the antineutrino channels $\bar{\nu}_\mu \rightarrow \bar{\nu}_e$ and $\bar{\nu}_\mu \rightarrow \bar{\nu}_\mu$, enabling the determination of the neutrino mass ordering.

The analysis of the generated MC samples is carried out in the following physics scenarios. We first examine the physics potential to exclude the wrong neutrino mass ordering at the ESSnuSB far detectors. The MC samples are then used to compute the sensitivity to the $\theta_{23}$ octant. We finally estimate the precisions at which the ESSnuSB setup can determine the leptonic mixing parameters $\theta_{23}$ and $\Delta m_{31}^2$ and provide the evolution of sensitivities to the neutrino mass ordering and the $\theta_{23}$ octant as functions of time. Both normal and inverted orderings are taken into account throughout this section.

\begin{figure}[!t]
        \center{\includegraphics[width=0.90\textwidth]{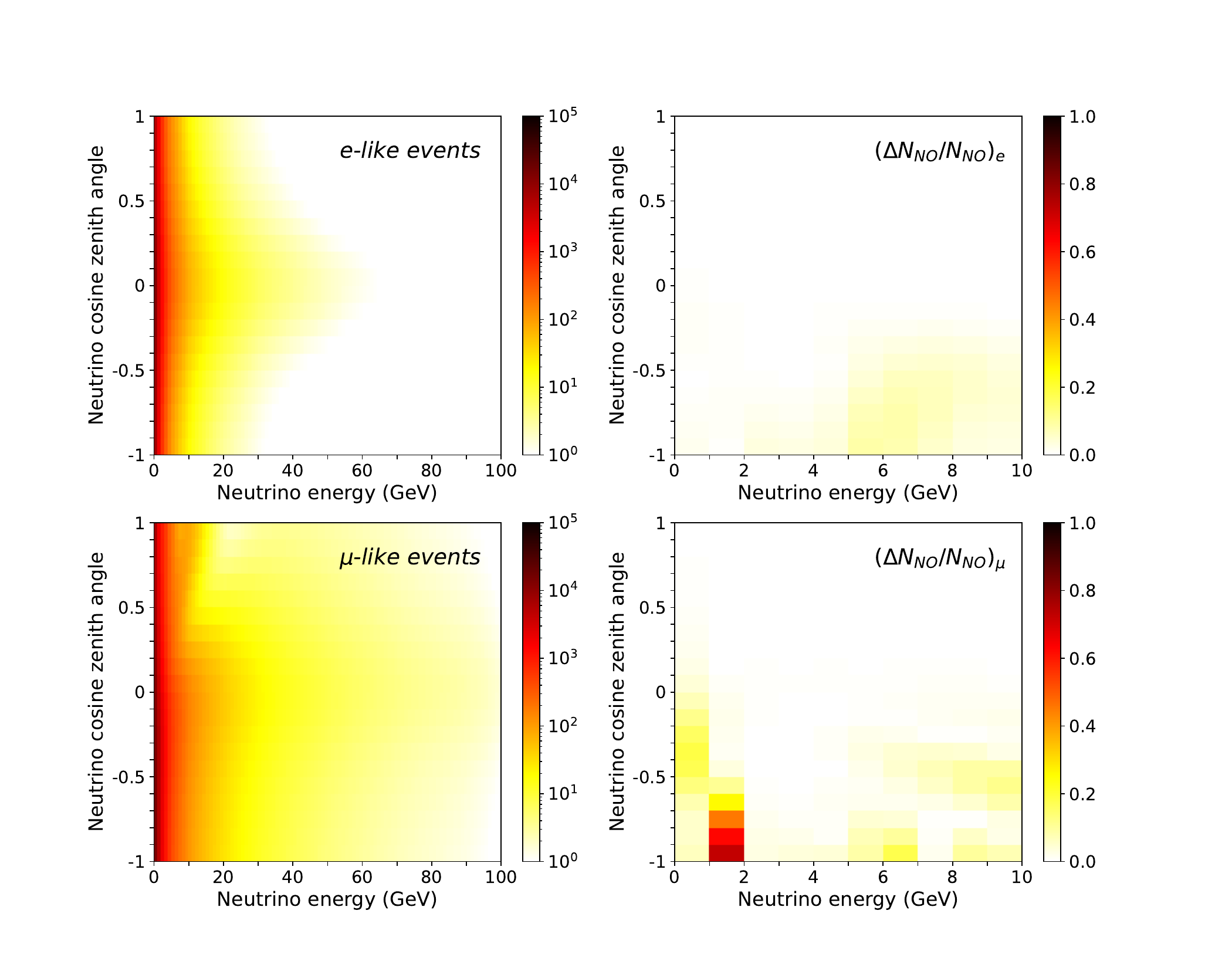}}
        \vspace{-0.65cm}
        \caption{Expected atmospheric neutrino spectrum in the ESSnuSB far detectors. Left panels show the total number of atmospheric neutrino events assuming normal neutrino mass ordering. Right panels present the relative differences in the atmospheric neutrino events under the normal ordering (NO) hypothesis.}
        \label{fig:events}
\end{figure}

The expected numbers of atmospheric neutrino events in the ESSnuSB far detectors are presented in Figure~\ref{fig:events}. The total number of atmospheric neutrino events corresponds to 5.4~Mt$\cdot$year exposure. The top-left and bottom-left panels of the figure show the total numbers of electron-like and muon-like events, respectively. The events have been binned for neutrino energies $E_\nu \in [0, 100]$~GeV and neutrino cosine zenith angles $\cos \theta_z \in [-1,1]$ with bin sizes $\Delta E_\nu = 1$~GeV and $\Delta \cos \theta_z = 0.1$. The top-right panel depicts the relative difference of the atmospheric neutrino events $\Delta N_{\rm NO} / N_{\rm NO} = |N_{\rm NO} - N_{\rm IO}| / N_{\rm NO}$ for the electron-like events, where $N_{\rm NO}$ represents the number of electron-like events for normal ordering and $N_{\rm IO}$ for inverted ordering. The bottom-right panel shows the same quantity for the muon-like events. The relative differences are shown for neutrino energies $E_\nu \in [0, 10]$~GeV. The determination of the $\theta_{23}$ octant and the precision measurements on $\sin^2 \theta_{23}$ and $\Delta m_{31}^2$ can be done by observing neutrino oscillations in any of the neutrino channels $\nu_\mu \rightarrow \nu_\mu$, $\nu_e \rightarrow \nu_\mu$, $\nu_\mu \rightarrow \nu_e$, as has been shown in equations~(\ref{eq:2.4a})--(\ref{eq:2.4c}). The sensitivities to these quantities are therefore proportional to the number of electron-like and muon-like events in the generated MC samples. On the other hand, the sensitivity to the neutrino mass ordering mainly arises from the difference in the number of neutrino events that are observed with the normal and inverted ordering hypotheses. The bottom-right panels of Figure~\ref{fig:events} shows that the relative difference between the two orderings is the most significant for the muon-like events in the neutrino energy bin $E_\nu \in [1, 2]$~GeV and neutrino cosine zenith angle bin $\cos \theta_z \in [-1, -0.9]$. The relative difference in this analysis bin is about 75\%. The next most significant contribution is found in the neutrino energy bin $E_\nu \in [0,1]$ and the neutrino cosine zenith angle bin $\cos \theta_z \in [-0.4,-0.3]$, where the relative difference in the muon-like events is about 20\%. For the electron-like sample, we find about 10\% relative difference in the neutrino energy bins $E_\nu \in [5,7]$~GeV and the neutrino cosine zenith angle bins $\cos\theta_z \in [-1,-0.8]$. The sensitivities furthermore depend on the systematic uncertainties\footnote{The sensitivity to the neutrino mass ordering arises mainly from the neutrino energy bin $E_\nu \in [5, 6]$~GeV, where the average difference between the charged lepton and neutrino zenith angles is about 9$^\circ$. We have explicitly checked that the contributions from lower neutrino energies are sub-dominant due to the effect of the systematic uncertainties.}.

\begin{figure}[!t]
        \center{\includegraphics[width=0.90\textwidth]{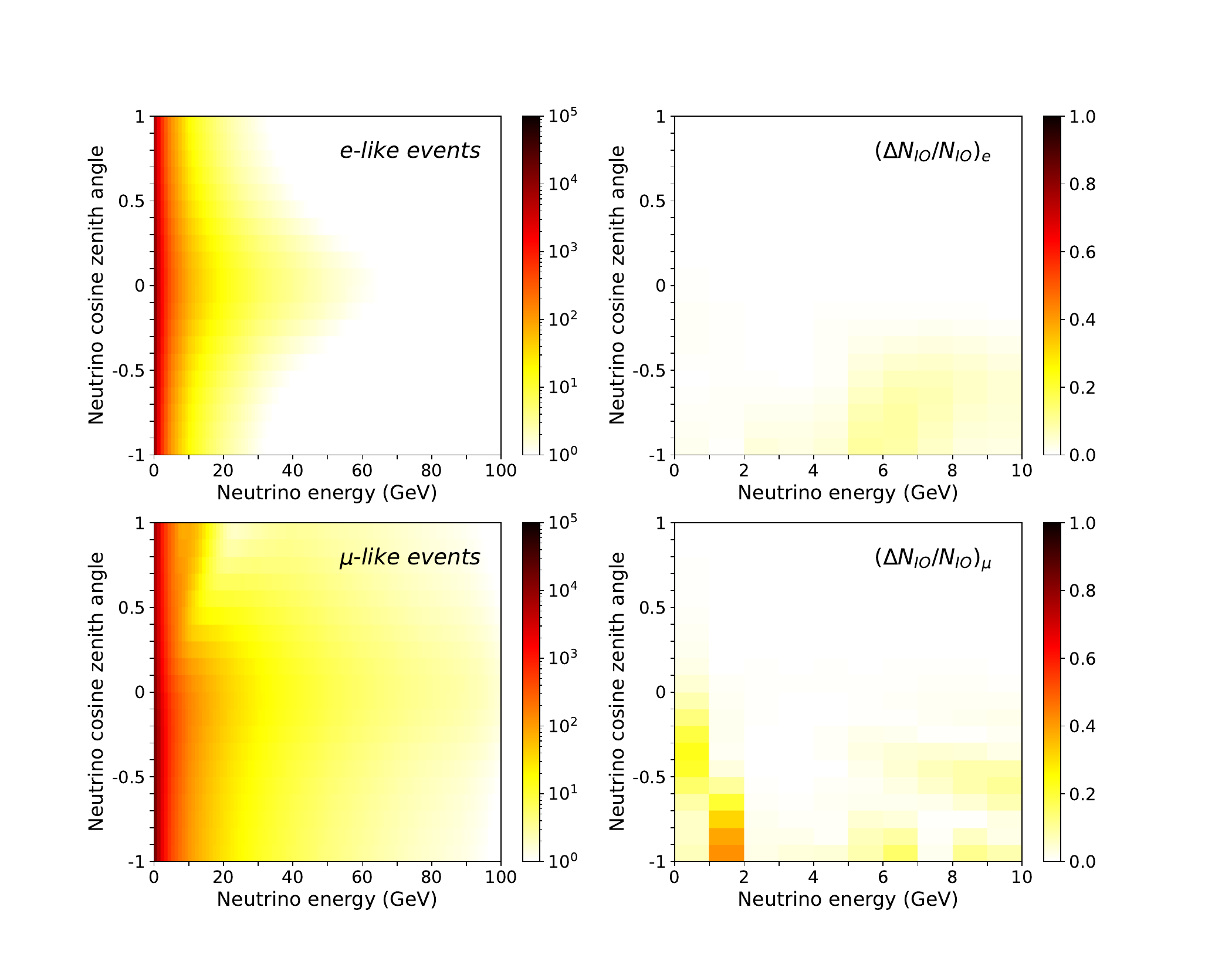}}
        \vspace{-0.65cm}
        \caption{Expected atmospheric neutrino spectrum in the ESSnuSB far detectors in the case of inverted ordering. Left panels present the total number of atmospheric neutrino events, whereas right panels show the relative differences. The true neutrino mass ordering is assumed to be the inverted ordering (IO).}
        \label{fig:eventsIO}
\end{figure}

The expected number of atmospheric neutrinos in the electron-like and muon-like samples and their relative differences are presented for the inverted ordering (IO) hypothesis in Figure~\ref{fig:eventsIO}. In this case, the expected numbers of electron-like and muon-like events are very similar to those presented for the NO hypothesis in Figure~\ref{fig:events}. The relative differences, which are defined as $\Delta N_{\rm IO} / N_{\rm IO} = |N_{\rm IO} - N_{\rm NO}| / N_{\rm IO}$ for the IO hypothesis, show that the contribution from the neutrino energy bin $E_\nu \in [1,2]$~GeV and the neutrino cosine zenith angle bin $\cos\theta_z \in [-1, -0.9]$ in the muon-like sample is considerably lower than the relative difference obtained in the NO hypothesis. The relative difference in these neutrino energy and neutrino cosine zenith angle bins for the muon-like sample is approximately 40\% for the IO hypothesis. For the electron-like sample, the relative differences are similar with respect to both IO and NO.

\begin{figure}[!t]
        \center{\includegraphics[width=1.00\textwidth]{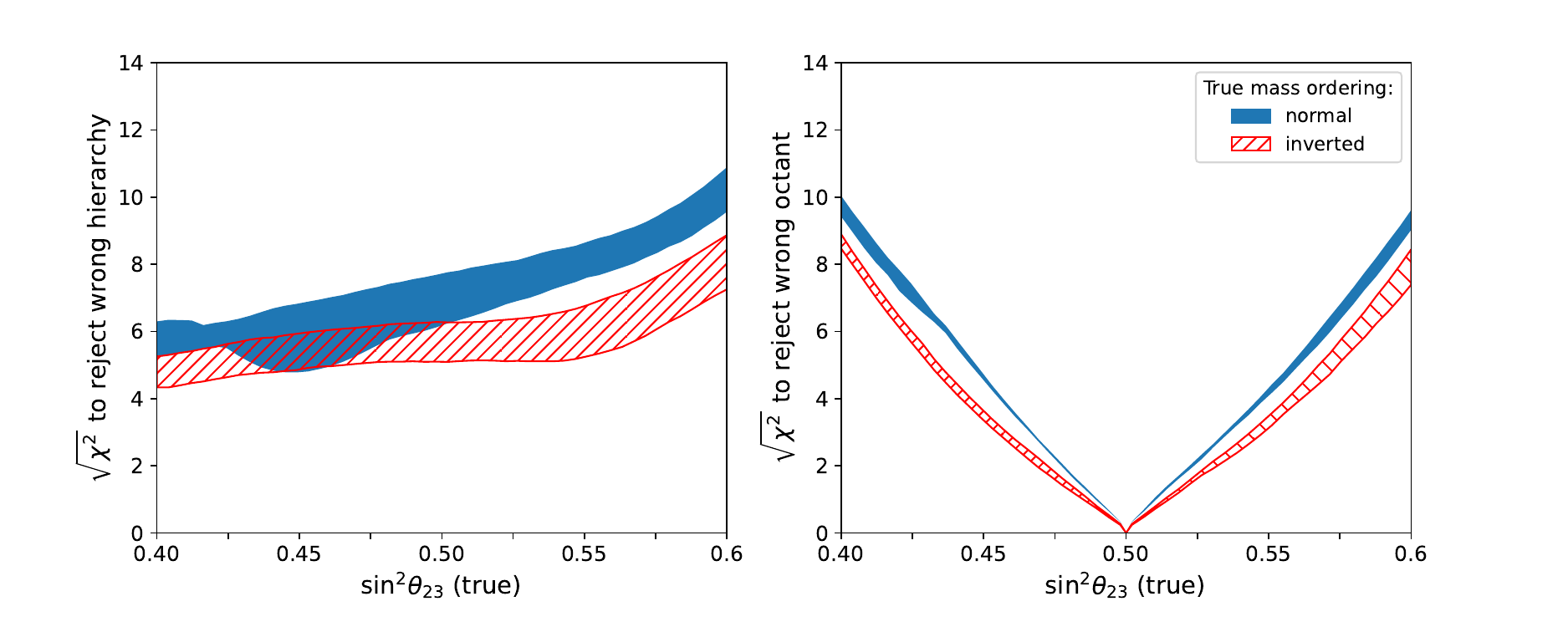}}
        \vspace{-0.5cm}
        \caption{Mass ordering and octant sensitivities in the ESSnuSB far detector with atmospheric neutrinos. Sensitivities are presented as functions of the true value of $\sin^2 \theta_{23}$, while assuming either normal ordering (blue bands) or inverted ordering (red hashes) as the true neutrino mass ordering. The line widths arise from varying the true value of $\delta_{CP}$.}
        \label{fig:sensitivities}
\end{figure}

The sensitivities to probe the neutrino mass ordering and the $\theta_{23}$ octant with the ESSnuSB setup are shown in Figure~\ref{fig:sensitivities}. The left panel shows the statistical significance at which the wrong neutrino mass ordering can be ruled out as a function of $\sin^2 \theta_{23}$ in the case when the true mass ordering is normal ordering (blue bands) and inverted ordering (red hashes), respectively. Therefore, the sensitivity to rule out the inverted ordering is shown by the blue band, whilst the sensitivity to rule out the normal ordering is shown by the red hash. The results are shown for the true values $\sin^2 \theta_{23} \in [0.4, 0.6]$. The right panel shows the corresponding sensitivity to rule out the wrong $\theta_{23}$ octant. The mass ordering sensitivities are given as the number of standard deviations $N_\sigma = \sqrt{\chi^2}$, which are computed by minimizing the $\chi^2$ function over the neutrino oscillation parameter values that are consistent with the wrong mass ordering. In other words, the true data is obtained under normal ordering and fitted data under inverted ordering when the true mass ordering is normal, and the other way round when the true mass ordering is inverted. Similarly, the $\theta_{23}$ octant sensitivities are obtained by minimizing over the parameter values that conform with the wrong $\theta_{23}$ octant. The band widths correspond to the dependence on $\delta_{CP}$, which is varied over the values $\delta_{CP} = 0, \pi/2, \pi$ and $3\pi/2$. ESSnuSB can be expected to determine the neutrino mass ordering by $4.8\sigma$--$10.9\sigma$~CL for normal ordering and $4.3\sigma$--$8.9\sigma$~CL for inverted ordering. The sensitivity to the neutrino mass ordering generally increases as a function of $\sin^2 \theta_{23}$, with a local minimum at $\sin^2\theta_{23} = 0.55$ for inverted ordering. The $CP$ phase $\delta_{CP}$ has a non-negligible role in the determination of the neutrino mass ordering. If the true neutrino mass ordering is the normal ordering, the variation in the neutrino mass ordering sensitivity is the largest at $\sin^2 \theta_{23} \simeq 0.45$ and lowest at $\sin^2 \theta_{23} \simeq 0.42$, where varying $\delta_{CP}$ causes the sensitivities to change by $2.0\sigma$~CL and $0.7\sigma$~CL, respectively. If the true neutrino mass ordering is the inverted ordering, the variation is about $1\sigma$~CL regardless of the $\sin^2 \theta_{23}$ value. The $CP$ phase $\delta_{CP}$ has conversely smaller effect on the determination of the octant of $\theta_{23}$, where the variation is less than $0.5\sigma$~CL for normal ordering and less than $1.1\sigma$~CL for inverted ordering. Increasing the number of scan points for $\delta_{CP}$ may affect the band widths, however, we do not expect the changes to be significant.

Atmospheric neutrino oscillations can provide competitive sensitivities for both the neutrino mass ordering and the $\theta_{23}$ octant determinations. The sensitivity to reject the inverted ordering for the true value $\sin^2 \theta_{23} = 0.451$ with the atmospheric neutrino oscillation data for ESSnuSB is about $4.8\sigma$--$6.8\sigma $~CL and normal ordering $4.9\sigma$--$6.0\sigma$~CL, as indicated by the blue band and the red hash, respectively. The sensitivity to reject the high octant solution $\sin^2 \theta_{23} > 0.50$ for the true value $\sin^2 \theta_{23} = 0.451$ is approximately $4.4\sigma$--$4.5\sigma$~CL for normal ordering and $3.3\sigma$--$3.5\sigma$~CL for inverted ordering. We remark that the zenith angle dependence constitutes the most significant systematic uncertainty in the determination of neutrino mass ordering. The other four systematic uncertainties also yield notable effects. For the true values $\sin^2\theta_{23} = 0.451$ and $\delta_{CP} = 1.29\pi$, for example, the systematic uncertainty pertaining to the zenith angle dependence leads to about $0.4\sigma$~CL reduction in the mass ordering sensitivity for NO, whereas the uncertainties related to the flux normalization, cross-section normalization and detector errors account $0.2\sigma$~CL and the energy tilt error $0.1\sigma$~CL, respectively. The total reduction due to the systematic uncertainties is about $0.7\sigma$~CL.

\begin{figure}[!t]
        \center{\includegraphics[width=1.00\textwidth]{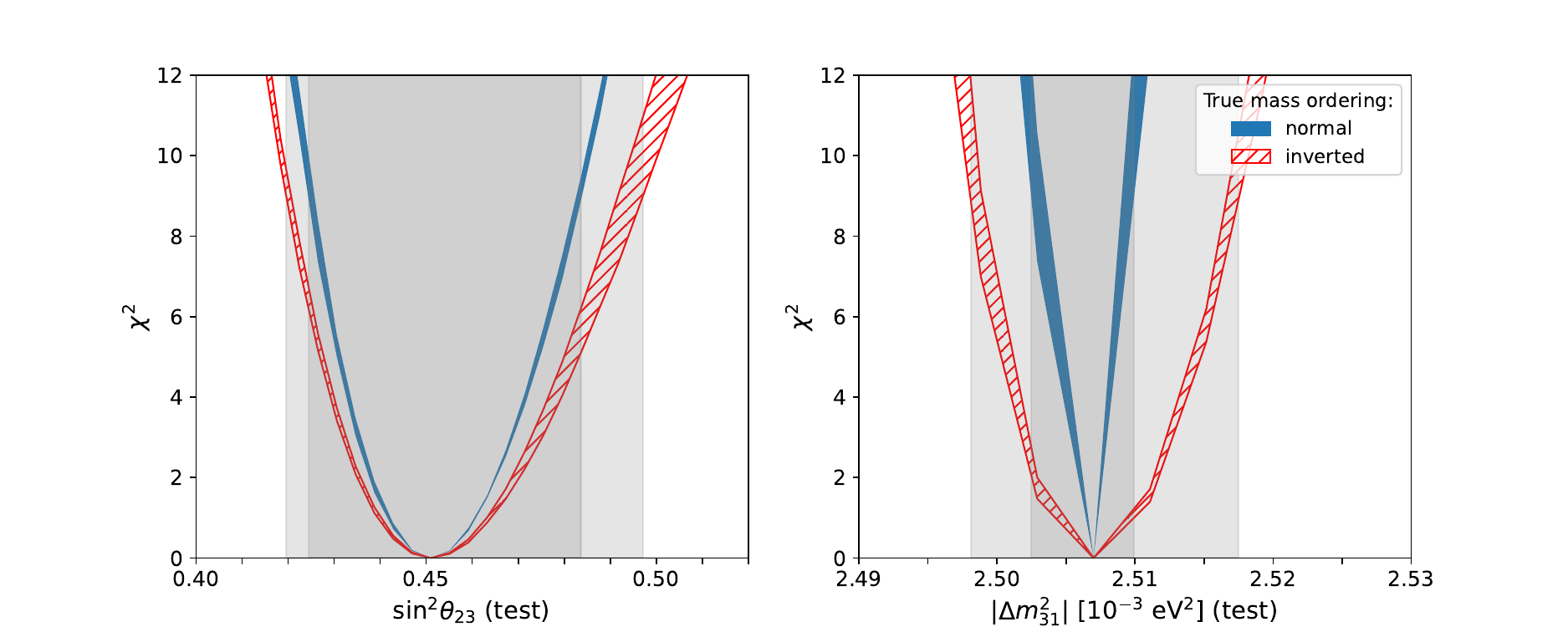}}
        \caption{Precision to $\sin^2 \theta_{23}$ and $\Delta m_{31}^2$ in ESSnuSB far detector by measuring atmospheric neutrino oscillations. $CP$ phase has been varied over $\delta_{CP} \in [0, 2\pi)$. Shaded areas indicate the allowed values of $\sin^2 \theta_{23}$ and $|\Delta m_{31}^2|$ at 3$\sigma$~CL for normal ordering (dark grey) and inverted ordering (light grey), respectively.}
        \label{fig:precisions}
\end{figure}

Atmospheric neutrino oscillations can be used to constrain the values of the leptonic mixing parameters $\theta_{23}$ and $\Delta m_{31}^2$. The precisions on these mixing parameters are illustrated in Figure~\ref{fig:precisions}. The one-dimensional $\chi^2$ distributions are presented as functions of $\sin^2 \theta_{23}$ (left panel) and $|\Delta m_{31}^2|$ (right panel) assuming the neutrino mass ordering to follow either normal ordering (blue bands) or inverted ordering (red hashes). In both panels, the true values of the relevant mixing parameters are assumed to be $\sin^2\theta_{23} = 0.451$ and $|\Delta m_{31}^2| = 2.507\times10^{-3}$~eV$^2$, whereas $\delta_{CP}$ is varied over its allowed values $\delta_{CP} \in [0, 2\pi)$. For convenience, the allowed $3\sigma$ CL ranges for ESSnuSB far detectors are shown with the dark grey areas for the case where the true neutrino mass ordering is the normal ordering and light grey areas for the case where the true neutrino mass ordering is the inverted ordering, respectively. The dark grey and light grey areas correspond to the $\sin^2 \theta_{23}$ and $|\Delta m_{31}^2|$ values where the lower edges of the blue bands and red hashes coincide with $\chi^2 = 9$. The lower edges of the grey areas therefore indicate the values of $\sin^2 \theta_{23}$ and $|\Delta m_{31}^2|$ that are allowed by $3\sigma$~CL using the atmospheric neutrino data from the ESSnuSB far detectors. As before, the results correspond to 5.4~Mt$\cdot$year total exposure. The mixing angle $\theta_{23}$ can be constrained to $0.424 < \sin^2\theta_{23} < 0.484$ for normal ordering and $0.419 < \sin^2\theta_{23} < 0.498$ for inverted ordering, whereas the magnitude of the mass-squared difference $\Delta m_{31}^2$ can be restricted to $2.502 \times 10^{-3}~\text{eV}^2 < |\Delta m_{31}^2| < 2.510 \times 10^{-3}~\text{eV}^2$ for normal ordering and $2.498 \times 10^{-3}~\text{eV}^2 < |\Delta m_{31}^2| < 2.518 \times 10^{-3}~\text{eV}^2$ for inverted ordering, respectively. As can be observed from the results, the effect of the $\delta_{CP}$ variation is relatively small both in the $\sin^2 \theta_{23}$ and the $\Delta m_{31}^2$ resolutions.

\begin{figure}[!t]
        \center{\includegraphics[width=1.00\textwidth]{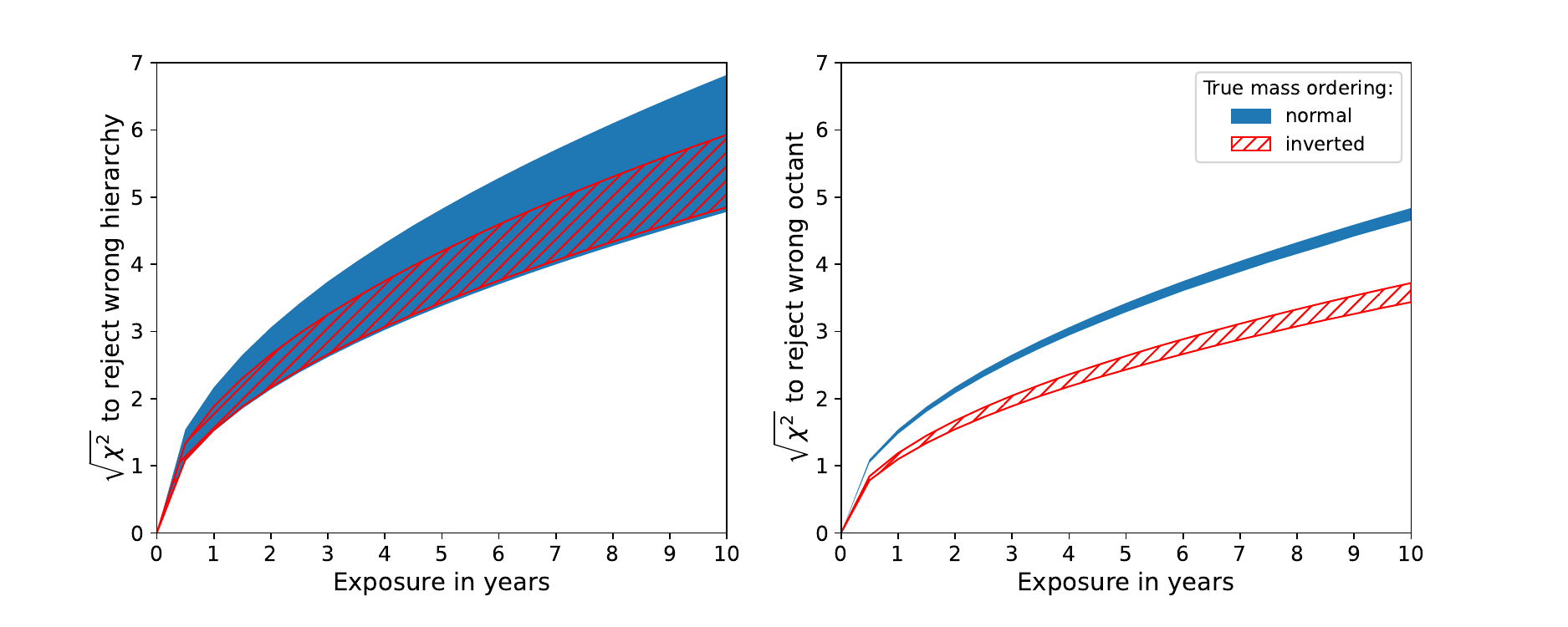}}
        \vspace{-0.5cm}
        \caption{Time dependence for the mass ordering and octant sensitivities in ESSnuSB far detector. The sensitivities are presented as a function of time which the ESSnuSB far detector is able to detect atmospheric neutrinos. The sensitivities are presented for both assuming normal and inverted orderings, whereas the line widths arise from the uncertainty on $\delta_{CP}$.}
        \label{fig:exposures}
\end{figure}

Figure~\ref{fig:exposures} presents the sensitivities to the neutrino mass ordering and the $\theta_{23}$ octant as functions of time. As before, the true values for the oscillation parameters are provided in Table~\ref{tab:scanrange} and the sign of $\Delta m_{31}^2$ is fixed according to the selected true mass ordering. For this choice of the neutrino oscillation parameter values, the sensitivities to the neutrino mass ordering overlap for normal ordering and inverted ordering. The corresponding sensitivities for the $\theta_{23}$ octant determination on the other hand are higher for normal ordering than for inverted ordering. The effect of the $\delta_{CP}$ variation is also lower in the $\theta_{23}$ octant determination. Figure~\ref{fig:exposures} shows that the wrong mass ordering can be ruled out at $3\sigma$ CL after about 4 years of data taking regardless of the true ordering, whereas the $5\sigma$ CL milestone can be reached for the majority of the considered $\delta_{CP}$ values. As before, the widths of both sensitivity bands correspond to the uncertainty on $\delta_{CP}$. The wrong $\theta_{23}$ octant on the other hand can be ruled out by $3\sigma$ CL after 4 years for normal ordering and 8 years for inverted ordering. 

\section{\label{sec:summary}Summary}

This work presents a study on atmospheric neutrino oscillations at the ESSnuSB far detector facility. The ESSnuSB project proposes a megaton-scale Water Cherenkov neutrino detector to observe neutrinos from the European Spallation Source. In addition to detecting neutrinos from the accelerator facility, the ESSnuSB far detectors would be capable of observing neutrinos from non-beam sources. Atmospheric neutrinos present an excellent opportunity to study neutrino oscillations with long-baseline lengths and strong matter effects, therefore complementing the physics program of the ESSnuSB project.

In the present work, we investigated the physics prospects of observing atmospheric neutrino oscillations at ESSnuSB in the standard three-flavor oscillation paradigm. The expected experimental sensitivities were examined for the determination of the neutrino mass ordering, the discovery of the $\theta_{23}$ octant and the precision measurements on $\theta_{23}$ and $\Delta m_{31}^2$. It is found that ESSnuSB is able to determine the correct neutrino mass ordering at $3\sigma$~CL after 4 years and $5\sigma$~CL after 10 years of data taking when the value of $\delta_{CP}$ is not known, regardless of the mass ordering. It is also shown that ESSnuSB would be able to determine the $\theta_{23}$ octant at $3\sigma$~CL after 4 years if the neutrino mass ordering is normal ordering and 8 years if it is inverted ordering. The atmospheric neutrino data collected by the ESSnuSB far detectors could also provide individual constraints on the values of $\theta_{23}$ and $\Delta m_{31}^2$. The sensitivities derived in this work are complementary to the beam-based long-baseline neutrino oscillation program for ESSnuSB.

\section*{Acknowledgements}
Funded by the European Union, Project 101094628. Views and opinions expressed are however those of the author(s) only and do not necessarily reflect those of the European Union. Neither the European Union nor the granting authority can be held responsible for them. The numerical calculations presented in this work were carried out in part with resources provided by the National Academic Infrastructure for Supercomputing in Sweden (NAISS), which is partially funded by the Swedish Research Council through grant agreement no. 2022-06725. The authors are also grateful for the computing resources that are provided by the Division of Condensed Matter Theory at KTH Royal Institute of Technology.

\bibliography{references}
\end{document}